\documentclass{aastex}
\usepackage{emulateapj5,apjfonts}

\usepackage{psfig}

\shorttitle{X-ray absorption in the high-$z$ quasar RX\,J1028.6-0844}
\shortauthors{Yuan W. et al.}

\def\ergse{${\rm erg\,s^{-1}\,cm^{-2}}$}
\def\ulum{${\rm erg\,s^{-1}}$ }
\def\ulume{${\rm erg\,s^{-1}}$}
\def\nh{$N_{\rm H}$ }
\def\nhe{$N_{\rm H}$}
\def\nhi{$N_{\rm H~I}$ }
\def\nhie{$N_{\rm H~I}$}
\def\nhex{$N_{\rm H}^{\rm exc}$ }
\def\nhexe{$N_{\rm H}^{\rm exc}$}
\def\unh{${\rm cm^{-2}}$ }
\def\unhe{${\rm cm^{-2}}$}
\def\abd{$A_{\rm Z}$ }
\def\abde{$A_{\rm Z}$}
\def\abdsol{${A_{\rm Z \sun}}$ }
\def\abdsole{${A_{\rm Z \sun}}$}
\def\alpox{$\alpha_{\rm ox}$ }
\def\alpoxe{$\alpha_{\rm ox}$}
\def\alprx{$\alpha_{\rm rx}$ }
\def\alprxe{$\alpha_{\rm rx}$}
\def\alpro{$\alpha_{\rm ro}$ }
\def\alproe{$\alpha_{\rm ro}$}
\def\gnh{$N_{\rm H}^{\rm Gal}$ }
\def\gnhe{$N_{\rm H}^{\rm Gal}$}
\def\src{RX\,J1028.6-0844 }
\def\srce{RX\,J1028.6-0844}
\def\gb{GB~1428+4217 }

\def\hi{\rm H~I }
\def\hie{\rm H~I}

\begin{document}

\title{Evidence of thick obscuring matter revealed in the X-ray
spectrum of the $z$=4.28 quasar RX\,J1028.6-0844}

\author{W. Yuan \altaffilmark{1}, M. Matsuoka \altaffilmark{1}, 
T. Wang \altaffilmark{2}, S. Ueno \altaffilmark{1}
H. Kubo \altaffilmark{3}, T. Mihara \altaffilmark{4}}

\altaffiltext{1}{Space Astrophysics group,
National Space Development Agency of Japan ({\it NASDA}),
Tsukuba Space Center, 2-1-1 Sengen, Tsukuba, Ibaraki 305, Japan
Email (YW): ywm@oasis.tksc.nasda.go.jp}
\altaffiltext{2}{Center for Astrophysics, University of Science and
Technology of China, Anhui, 230026, China}
\altaffiltext{3}{Department of Physics, Tokyo Institute of Technology, 2-12-1,
Ookayama, Meguro, Tokyo, 152-8551, Japan}
\altaffiltext{4}{The Institute of Physical and Chemical Research ({\it
RIKEN}), 2-1, Hirosawa, Wako, Saitama, 351-0198, Japan}

\begin{abstract}
We report the discovery of an unambiguous, substantial low-energy cutoff 
in the broad band X-ray spectrum of the radio quasar \src at a redshift of 4.276
obtained with the ASCA satellite,
which we preferably explained as indication of excess X-ray absorption.
The equivalent hydrogen column density of the absorbing matter, 
depending on the redshift and metallicity, 
ranges from $2.5~10^{21}$\,\unh for local absorption up to
$2.1~10^{23}$\,\unh (solar metallicity) or 
$1.6~10^{24}$\,\unh (10\% solar metallicity) 
for absorption at the quasar redshift.
Such a value is among the largest found for high-redshift radio quasars. 
The absorption, if interpreted as being produced close to the quasar,
may indicate the presence of a remarkably large amount of 
obscuring matter in the quasar environment in the early universe.  
Implications of the result for the possible origins of the absorbing matter
are discussed, concerning especially galactic intervening matter, 
cool intracluster gas, and ambient medium around the quasar jet. 
The quasar itself has an enormous apparent luminosity 
of at least about $2.6~10^{47}$\,\ulum
($H_0=50$\,km\,s$^{-1}$\,Mpc$^{-1}$, $q_0=0.5$) 
and a power law photon index of $1.67(^{+0.07}_{-0.04})$ 
in the 2--50\,keV band in the source rest frame.
\end{abstract}


\section{Introduction}

Photoelectric absorption in the soft X-ray spectra of quasars provides
unique probes of circum-source matter and quasar evolution.
There is growing evidence that radio-loud quasars at high redshifts ($z>2$)
are commonly obscured by opaque matter---the X-ray spectra
flatten toward low energies and this is attributed to 
absorption of the soft X-ray photons
in excess of that due to Galactic interstellar medium, 
known as the excess absorption 
\citep{wilkes92,elvis94,serlem94,siebert96,cappi97,reeves,fiore98}.
It is not known what cause such absorption, nor where they occur. 
Both, cosmologically intervening and quasar-associated materials are
proposed as absorber candidates; 
and the latter is favored by statistical arguments  
\citep{bechtold94,oflah97,fiore98,yuan99} 
for high-$z$ radio-quiet quasars showing no excess absorption, 
and by tentative indications for variability of the absorption 
(e.g.\ Schartel et al.\ 1997, Cappi et al.\ 1997). 
This picture has been established based mainly on observations of quasars
in the redshift range of 2--4. 
The highest redshifts up to which the excess absorption was reported 
in these observations are of
PKS\,2126-158 at $z=3.3$ \citep{elvis94} and of
S5\,0014+81 at $z=3.4$ \citep{cappi97},
and probably of Q1745+624 at $z=3.9$ \citep{kubo97}.
 
At even higher redshifts ($z>4$), only two quasars, both radio-loud,
were observed with fair broad band X-ray spectroscopy, 
1508+5714 at $z$=4.3 \citep{moran97} 
and GB~1428+4217 at $z$=4.7 \citep{fabian98}. 
While the ASCA spectra of both objects do not require excess absorption,
GB~1428+4217  was recently reported to have excess absorption 
with a ROSAT PSPC observation
(a hydrogen column density of $1.52(\pm 0.28)~10^{22}$\,\unhe, 
assuming the absorber associated with the quasar; Boller et al.\ 2000).
Another interesting and probably related aspect is that 
both objects show blazar properties, 
i.e.\ radio and X-ray emission dominated by 
relativistically beamed components from the jets.  
Though two objects are too few for any statistical inference,
it is intriguing to wonder whether this behavior is typical 
at very high redshifts or merely selection effects.

In this paper, we report an ASCA observation of the $z>4$ quasar
\srce, as part of our X-ray high-$z$ quasar program \citep{matsuoka99}.
The quasar shows an unambiguous, strong low-energy cutoff
in its X-ray spectrum.    
The X-ray source \src was discovered in the ROSAT All-sky Survey (RASS)
and identified with a quasar at $z=4.276$ by \citet{zickg97}. 
Its X-ray colors in the ROSAT energy band 
indicated an extremely hard spectrum, thus, strong X-ray absorption
if the `intrinsic' spectrum is of typical radio-loud quasars \citep{zickg97}. 
The object is also associated with a radio source, PKS\,B1026-084 
(Otrupcek \& Wright 1991), which has a flux density of 220\,mJy at 5\,GHz
and a flat radio spectrum ($\alpha=0.3$, $S\propto \nu^{-\alpha}$);
hence it is a flat-spectrum radio-loud quasar (FSRQ), 
and has an intense 5\,GHz radio
luminosity\footnote{$H_0=50$\,km\,s$^{-1}$\,Mpc$^{-1}$ 
and $q_0=0.5$ are assumed throughout the paper.} 
$\nu L_{\nu}= 5~10^{44}$\,\ulume.
We present the observation and data reduction in Sect.\,2, and
spectral and temporal analyses in Sect.\,3 and 4, respectively. 
In Sect.\,5 the implications for the origins of the excess absorption
are discussed. A summary of the main results is given in Sect.\,6.
Errors quoted are of the 68\% confidence level (c.l.) throughout the paper 
unless mentioned otherwise.

\section{Observation and data reduction}

\src was observed with the ASCA satellite \citep{tanaka94} on Nov.\ 25,
1999 and the duration lasted 67 hours.
The Solid-state Imaging Spectrometers (SIS) were operated 
in the 1-CCD faint mode
and the Gas Imaging Spectrometers (GIS) in pulse height mode.
For SIS the `bright~2' mode data (grade 0, 2, 3, 4) were used,
which were converted from the original faint mode data.
Corrections for the dark frame error and the echo effect were applied. 
The data reduction was performed in the standard way by using the FTOOLS (v.4.2)
utilities. Hot and flickering pixels were removed from the SIS data.
To improve the reliability of the results,
we chose conservative data screening criteria,
which are listed in Table\,\ref{tab:observation}.

The source counts were extracted from circular regions of 3.5 and 5 arcmin
radii for SIS and GIS, respectively.
The backgrounds were determined in two ways and compared with each other:
from the blank-sky observations using the same region of the detector, 
and from source-free regions in the field of view of the same observation 
(local). 
For the latter, for each of the GIS detectors two local off-source regions 
were used to extract background events, which were chosen so as to have
the same radii and off-axis angles as the source region;
for SIS, the background events were extracted from the 
whole CCD chip with a circular region of 5\,arcmin
radius around the source excluded.
In addition to closely matching the screening criteria,
the local backgrounds have the advantages over the blank-sky backgrounds
regarding the uncorrected decreasing of the low-energy efficiency for SIS
and the variability of the internal background for GIS.
This is especially true considering the long time span between 
the blank-sky and the current observations. 
We thus mainly used the local backgrounds to derive results 
and used the blank-sky backgrounds for comparisons.
The resulting effective exposures and extracted source counts are
given in Table\,\ref{tab:observation}.

Since the two GIS have identical response matrices and 
the two source spectra agree well with each other,
we combined them into one single GIS spectrum\footnote{The 
counting errors in each bin were assigned as pure Poissonian errors
of the summed counts.}.
All the spectra were re-binned to have at least 30 counts in each energy bin. 
We used the most updated releases of the response matrices available so far
(V4.0 GIS RMF gis2(3)v4\_0.rmf, V1.1 SIS RMF and the calibration file 
sisph2pi\_110397.fits) in spectral fit. 

\section{Spectral analysis}
\label{excsabs}

It is known that the responses of the two SIS suffer from 
calibration uncertainties, 
which are the most problematic at around 0.6\,keV
(ASCA Guest Observer Facility web page). 
To minimize the possible systematic error introduced in the results,
we ignored the SIS data below 0.7\,keV in spectral fit,
but compared them with the fitted models.
The energy ranges used for spectral fits are 0.7--9.5\,keV for both
SIS and GIS.
We first fit the spectra with a simple power law model modified by 
photoelectric absorption assuming an absorber in the observer frame
(local absorption, Sect.\,\ref{excsabs_loc}) and at a redshift  
(redshifted absorption, Sect.\,\ref{excsabs_red}), respectively,
in addition to the Galactic absorption.
Then we test other spectral models for the continuum in Sect.\,\ref{fit_nonpl}.
We consider the impact of the excluded low-energy data 
on the obtained results in Sect.\,\ref{certainty}.

\subsection{Power law with local absorption model}
\label{excsabs_loc}

\begin{figure}
\psfig{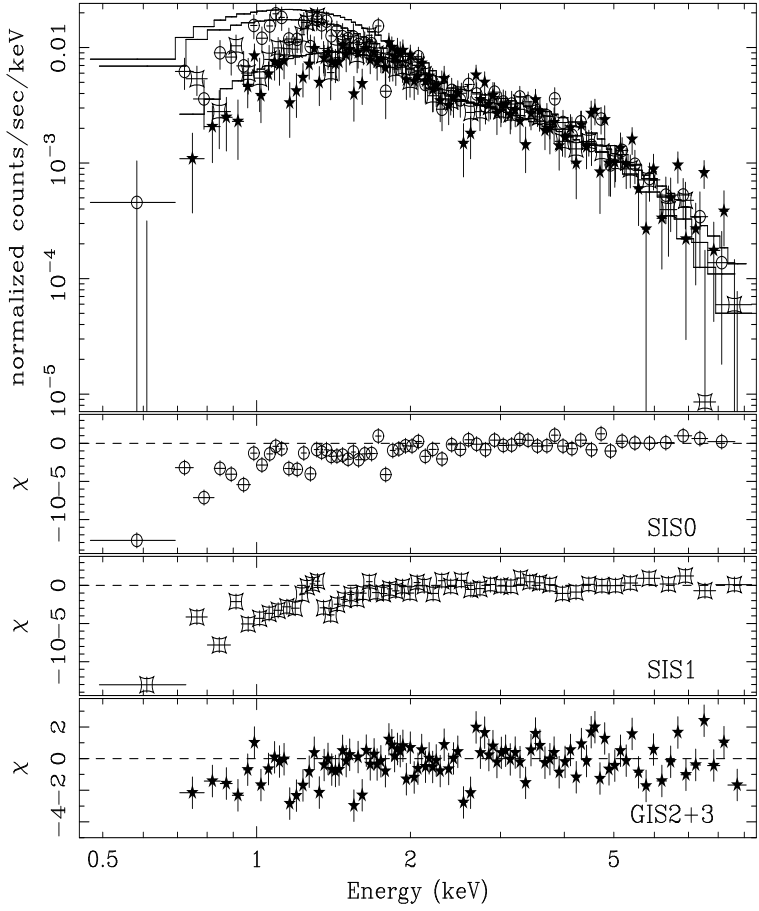}
\caption[]{{\bf Upper panel}: Joint spectral fit of 
a power law with fixed Galactic absorption
to the SIS0 (circles), SIS1 (squares), and the combined GIS2+GIS3 (stars) data 
in the restricted 2.5--9.5\,keV energy band (with the normalizations untied).
The model values (convolved with the detector responses) are represented
by lines from upper to lower (at low energies) for SIS0, SIS1, and GIS2+3,
respectively. 
{\bf Lower panel}: 
Residuals showing deviations of the data from the
extrapolation of the fitted model
down to the low energy range for all the detectors, respectively.
}
\label{fig:pl-gnh-extrap}
\end{figure}

The Galactic column density in the source direction is
\gnh$=4.59~10^{20}\,\rm cm^{-2}$ (Dickey \& Lockman 1990).
A fit with a single power law model absorbed by `cold' gas 
(model {\em wabs}\footnote{For which the photoelectric
absorption cross sections from Morrison and McCammon (1983) are used.}
in {\em XSPEC} v.10) 
with the column density fixed at \gnh
gave a flat photon index $\Gamma = 1.34\pm0.03$
(joint fit of the SIS0, SIS1, and GIS spectra, see Table\,\ref{tab:spec_fits} 
for the fitted parameters).
Though the model is statistically acceptable, 
the data below 1\,keV fall systematically below the model. 
The deviations are more pronounced as shown in
Fig.\,\ref{fig:pl-gnh-extrap}, in which the above model 
was fitted to the spectra in the restricted
2.5--9.5\,keV band only (13--50\,keV in the quasar frame) 
and extrapolated down to 0.5\,keV. 
Such a low-energy cutoff cannot be explained by the known 
calibration uncertainty for SIS.
This is because, 
firstly, the GIS spectra, though being less sensitive in the low energy range,
show the same effect with similar amounts of deviations as the SIS spectra; 
secondly, the loss of the SIS efficiencies is quantified 
as up to about 20--40\% at 0.6\,keV
(ASCA Guest Observer Facility web page),
while in this case the deviations of the data from the model
are about a factor of two on average, for both SIS and GIS.

\begin{figure}
\psfig{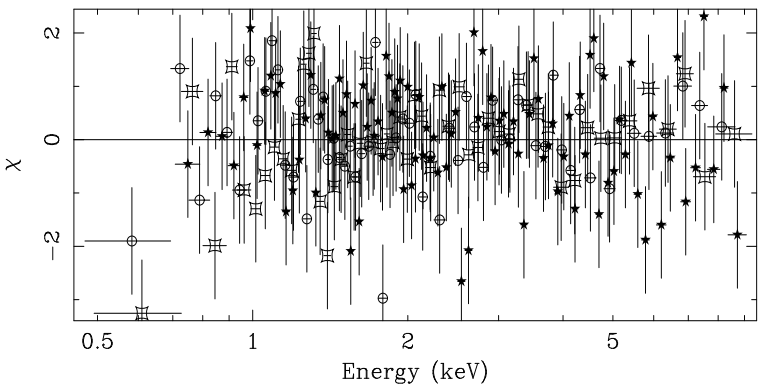}
\caption[]{Residuals for the best-fit model---a power law with 
free absorption column density to the joint SIS and GIS data.
The SIS data below 0.7\,keV are not used in the fitting but plotted for a comparison. 
The plot symbols are the same as in Fig.\ref{fig:pl-gnh-extrap}.}
\label{fig:best-fit}
\end{figure}

\begin{figure}
\psfig{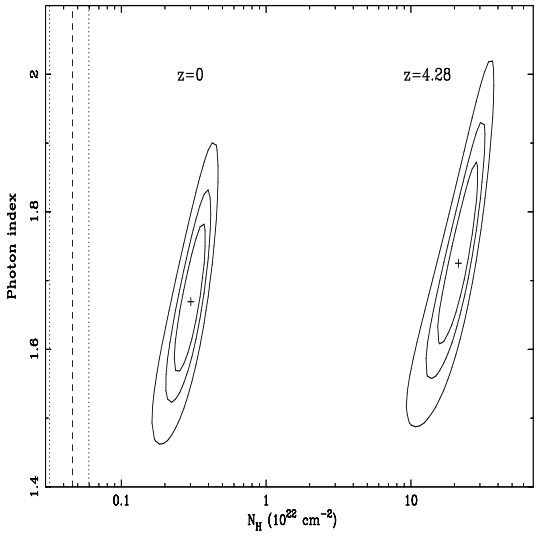}
\caption[]{Confidence contours 
of fitted absorption column density and photon index
for the total absorption at redshift $z$=0 and for the excess absorption
(with additional fixed Galactic absorption) 
assumed at the quasar redshift $z$=4.28, respectively.
The contours are at the 68\%, 90\%, and 99\% confidence, respectively, for two 
interesting parameters.
The elemental abundances are assumed to be the solar values.
The best-fit values are indicated by crosses. 
The Galactic column density and a conservative 30\% uncertainty
region are indicated by dashed and dotted lines, respectively. 
}
\label{fig:cont-nh-gam}
\end{figure}

Setting the absorption column density 
\nh as a free parameter improved the fits significantly, 
and yielded large \nh for both SIS and GIS.
The $\chi^2$ value was reduced by 
$\Delta\chi^2=$6, 18, and 34 for fitting the GIS, 
SIS0+SIS1, and the joint GIS+SIS spectra, respectively.
Thus, the addition of the excess absorption term 
is significant at $>99.99\%$ c.l.
For the joint GIS+SIS fit this gave 
$N_{\rm H}= 3.0(_{-0.4}^{+0.5})~10^{21}\,\rm cm^{-2}$, 
and a photon index $\Gamma = 1.67(^{+0.07}_{-0.04})$, which is typical of
FSRQ in the $\sim$1--10(20)\,keV band at low and medium redshifts 
(e.g.\ Siebert et al.\ 1996; Lawson and Turner 1997;
Reeves et al.\ 1997; Cappi et al.\ 1997; Brinkmann et al.\ 1997).
The residuals of the fit are shown in Fig.\,\ref{fig:best-fit}.
The SIS 0.5--0.7\,keV data, which were excluded from the spectral fitting, 
are also plotted; it is shown that 
the extrapolation of the model does not under-predict the count rates
in this band (though the real count rates could be somewhat higher because of 
the loss of the SIS efficiencies).
Fig.\,\ref{fig:cont-nh-gam} shows
the confidence contours for two interesting parameters
for the fitted \nh and $\Gamma$ (marked by $z=0$).
Absorption of the quasar X-rays 
in excess of that due to the Galactic column density is evident.
Using the background spectra derived from the blank-sky observations 
gave consistent results.

Due to the increasing degradation of the response,
the calibration for both SIS is uncertain below $\sim 1$\,keV;
this effect reportedly results in deficits of 
the observed counts at low energies
and, effectively, apparent `excess absorption'.
From the most recent quantification of this problem (Yaqoob et al.\ 2000),
the apparent excess \nh at the time of the observation  
was estimated as (using the 0.5--10\,keV data)
$\Delta N_{\rm H}\sim 0.7~10^{21}$\,\unh for SIS0 and
about $1.0~10^{21}$\,\unh for SIS1.
Fitting the {\em whole} 0.5--10\,keV SIS spectra yielded
\nh = $2.9(^{+0.7}_{-0.5})~10^{21}$\,\unh ($\Gamma=1.62$) for SIS0 and
      $4.5(^{+1.0}_{-0.7})~10^{21}$\,\unh ($\Gamma=1.74$) for SIS1;
the resulting excess absorption column densities are 
significantly larger than the estimated systematic uncertainty.
Thus, the observed low-energy cutoff is a true feature of the source spectrum,
not an artifact due to the calibration problem.
Very roughly, we estimated the `true' absorption column density 
by subtracting the systematic excess $\Delta N_{\rm H}$ from the fitted value,
as $N_{\rm H}-\Delta N_{\rm H}$.
We found that, for both SIS, the remaining column densities 
are very close to the fitted \nh values by using the 0.7--10\,keV spectra,  
for which the calibration uncertainty is thought to have less effect. 
Independently, the GIS spectrum gave a \nh
$3.0(^{+1.2}_{-1.0})~10^{21}$\,\unh in good agreement with that
fitted using the 0.7--10\,keV SIS spectra 
(see Table\,\ref{tab:spec_fits}), suggesting that
the SIS calibration systematic uncertainties are 
small and negligible in the 0.7--10\,keV band
in comparison to the statistical uncertainties.
We therefore regard the spectral parameters derived by using 
the 0.7--10\,keV SIS spectra, as performed throughout this work,
as being little affected by the calibration uncertainties,
and ignore this effect in the following analyses.

For the power law with Galactic absorption model,
the absorption corrected 2--10\,keV flux is $1.2~10^{-12}$\,\ergse,
corresponding to a luminosity of 
$6.2~10^{46}$ ($2.4~10^{47}$)\,\ulum 
in the 2--10\,(2--50)\,keV band in the quasar rest frame.
For the free absorption model,
the luminosity corrected for the {\em total} (Galactic + excess) absorption 
is $9.8~10^{46}$\,($2.6~10^{47}$)\,\ulum in the 2--10\,(2--50)\,keV band.
These values were obtained using the GIS data,
which are slightly higher than the SIS measurements by less than 10\%.
No iron K emission line was detected; 
the upper limit of the equivalent width for a narrow 
($\sigma=0.01$\,keV) line at 6.4\,keV 
is 195\,eV at the 90\% c.l. in the source rest frame.
We found no iron K-shell absorption edge feature for local absorption,
setting a 90\% upper limit on the optical depth as $\tau<0.79$ at 7.1\,keV.
 
\subsection{Power law with redshifted absorption}
\label{excsabs_red}

\subsubsection{Absorption at the quasar redshift}
Assuming the absorber is associated with the quasar, 
we fitted the spectra with models of a power law
modified by absorption occurring at $z=4.276$, and further
by the (fixed) Galactic absorption. 
For `cold' absorption models 
this results in the same $\chi^2$ with,  
but a larger excess absorption column density 
\nhex ($2.1~10^{23}$\,\unhe, see Table\,\ref{tab:spec_fits}) 
by two orders of magnitude than,
those for the above local absorption model.
This is because the absorbed photons have increased energies 
(by a factor of 1+$z$) and thus 
decreased absorption cross sections in the absorber frame. 
We over-plotted in Fig.\,\ref{fig:cont-nh-gam}
the confidence contours for the fitted \nhex and $\Gamma$ (marked by $z=4.28$).
It should be noted that we have assumed the solar elemental abundances 
for the absorbing gas.
The energy range of $E>0.7$\,keV in which the X-ray photons are absorbed 
in the ASCA bandpass corresponds to $E\ga 3.5$\,keV in the quasar rest frame; 
at such energies the cross section of photoelectric absorption is 
dominated by heavy elements, mainly O, Ne, Si, S, Ca, and Fe.
Since at high redshifts the metal abundances could be well sub-solar
(about 10\% or less, e.g.\ Lu et al.\ 1996; Pettini et al.\ 1997; 
Prochaska \& Wolfe 2000),
we assumed the metal abundances as \abd=0.1\,\abdsol and repeated the fit
(using {\it zvfeabs} model in {\em XSPEC}).
The resulting absorption column density becomes as high as 
\nhex$\sim 1.6~10^{24}$\,\unh (Table\,\ref{tab:spec_fits}). 
Since the true elemental abundances are unknown,
the determination of \nhex is subject to a large uncertainty, however.

We searched for any redshifted iron K absorption edge 
by fitting the edge separately from the redshifted absorption model. 
The best-fit edge energy is $7.2(^{+0.6}_{-1.2})$\,keV; 
the optical depth at the threshold energy is $\tau=0.15(^{+0.12}_{-0.10})$,
which is consistent the prediction from the fitted \nhex 
($\tau=0.26$ for \abde=\abdsole).
However, given the reduction in $\chi^2$ of $\Delta\chi^2=1.3$ only,
the detection of the iron K edge cannot be justified.
The non-detection of a significant iron edge gave no 
determination of the redshift of the absorber, unfortunately.

It is possible that the absorber is photoionized 
by the intense radiation from the quasar. 
A photoionized absorption model ({\em absori} in {\em XSPEC}) was fitted 
to the spectrum, in which the power law X-ray emission of the quasar
was taken as the ionizing continuum.
The result is inconclusive, however. 
The fit converged at neutral absorption with
the best-fit ionization parameter 
(as defined in Done et al.\ 1992) $\xi \simeq 0$, 
while ionization of the gas to some degree is allowed, 
$\xi< 2$ (20) at the 90\% (99\%) c.l.
For sub-solar metal abundances the allowed $\xi$ range becomes lower,
$\xi<0.1$ at the 90\% c.l. for \abd=0.1\abdsole.
It should be noted that, however,
if the relative abundance of iron with respect to other heavy elements
is much lower than the solar value,
the ionization state is almost unconstrained,
though this seems to be an unlikely case.   
For instance, for \abd=\abdsol except $A_{\rm Fe}=0.1\,A_{\rm Fe \sun}$,
the 90\% confidence range for $\xi$ could be as large as 0--1000.
This is because the constraint on the ionization state comes largely from
the lack of the predicted K-edge features of highly ionized irons 
in 1.3--1.7\,keV in the observer frame.

\subsubsection{Redshift dependence}
\begin{figure}
\psfig{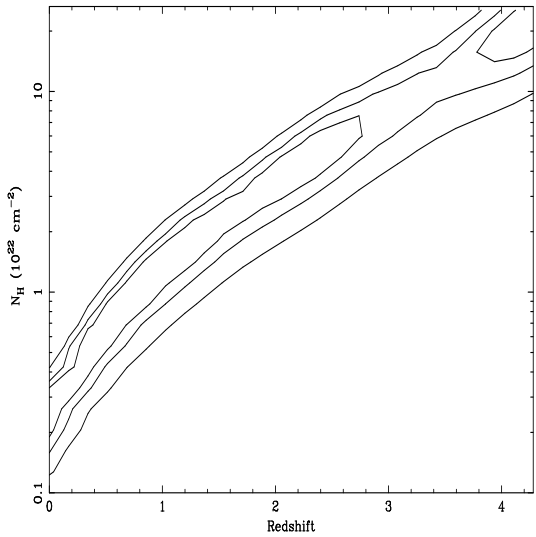}
\caption[]{Contours of allowed excess absorption \nhex and the redshift 
of the absorber for two interesting parameters at the 68\%, 90\%, and 99\% 
confidence levels, respectively. 
The elemental abundances are assumed to be the solar values.}
\label{fig:cont-nh-z}
\end{figure}

The unknown location of the absorber
introduces a large uncertainty in the determination of \nhexe. 
With the limited low-energy bandpass and energy resolution,
ASCA cannot distinguish between local and highly redshifted absorption models. 
This is because, firstly, these models could have similar spectral shapes
of the absorbed continuum in the ASCA band, 
and secondly, the photon statistics are generally
poor at low energies anyway, especially in the case of strong absorption. 
Leaving the redshift free yielded no statistically
significant range in the parameter space, i.e.\ similar  
goodness-of-fits for redshifts in the full range of 0--4.28,
though there exist a few shallow local minima in the $\chi^2$ distribution
(however, see Sect\,\ref{certainty} for the case when the 
0.5--0.7\,keV spectral data are included).
The dependence of the excess column density on the assumed redshift 
is shown in  Fig.\,\ref{fig:cont-nh-z}, in which the 
confidence contours are plotted on the \nhex--$z$ plane.
The required \nhex decreases down to $2.5~10^{21}$\,\unh for $z$=0,
as shown in Sect.~\ref{excsabs_loc}.
It is also shown that, at a given redshift, \nhex is relatively well constrained.
It should be noted that these results 
were obtained by assuming the solar elemental abundances;
if the metallicity is significantly sub-solar, 
the inferred \nhex value would increase substantially (see above).

\subsection{Other continuum models}
\label{fit_nonpl}

{\it Thermal bremsstrahlung model}---
We also fitted the spectrum with redshifted thermal bremsstrahlung 
plus local absorption models.
For absorption fixed at the Galactic \gnhe, 
a source frame temperature as high as $132(_{-27}^{+41})$\,keV is required.
Free fitting the absorption column density 
gave $T=55(_{-9}^{+12})$\,keV and, similarly,
a large \nh=$2.1(^{+0.4}_{-0.3})~10^{21}$\,\unhe.
The resulting goodness-of-fit is the same as that for a power law fit.
For such a high temperature, the thermal
bremsstrahlung spectrum is indistinguishable 
from a power law one in the ASCA band.

{\it Broken power law}---
A convex spectral shape, like what observed in some BL~Lac objects,
can result in apparent excess absorption
when fitted with a power law model.
Such spectra are usually parameterized with a broken power law model.
Fitting a broken power law with fixed Galactic absorption yielded
statistically acceptable fit, $\chi^2=190$ for 209 
degrees of freedom (d.o.f.);
in spite of one more free parameter,
the $\chi^2$ is somewhat worse than those for
the power law with excess absorption models. 
The resulting break energy of $9\pm 1$\,keV (in the quasar frame)
is higher than those found for the X-ray spectra of BL~Lac objects
(a few keV or less, e.g.\ Barr et al.\ 1988; Madejski et al.\ 1991).
Leaving the absorption \nh free improved the fits slightly
($\Delta \chi^2=3$), and yielded similar values for 
the high- and low-energy photon indices (Table\,\ref{tab:spec_fits}), 
and a \nh $\sim 2.8(\pm 0.7)~ 10^{21}$\,\unh
significantly larger than the Galactic value.
Such a model is effectively consistent, within the uncertainties of
the fitted parameters, with the absorbed power law models---the best-fit
models as shown above.  

\subsection{Effect of the low-energy end data: excess absorption?}
\label{certainty}

As shown above, the 0.7--9.5\,keV spectrum 
can be similarly well modeled by 
either the power law with (local or redshifted) excess absorption models
or the broken power law continuum with Galactic absorption. 
Since these models show increasing spectral divergence 
toward low energies, the omitted 0.5--0.7\,keV SIS data,
even in the presence of the calibration uncertainties, 
might be useful for discriminating between these models. 
Here we examine the effect of the excluded 0.5--0.7\,keV SIS data
(two bins after spectral binning) on the obtained results in two ways. 

Firstly, we extrapolated the above best fits for various models 
obtained with the 0.7--9.5\,keV spectrum
down to 0.5\,keV for SIS and compared 
the model values with the data.
For all the models, deficits of the observed count rate in these energy bins 
were always found for both SIS0 and SIS1, 
which might be a result of the above systematic effect, 
or mismatched models, or both.
We examined the consistency of the observed 0.5--0.7\,keV count rates 
with the current models by evaluating the {\it F}-statistic $F$, 
which basically measures the change in $\chi^2$ 
with the change of the degrees of freedom in fitting.  
(For a $F$ as large as exceeding the $F_{\rm cr}$ which corresponds to 
a certain probability level $P(F_{\rm cr})$, 
the chance probability under the model hypothesis
is as small as $P < P(F_{\rm cr})$).
For models of the power law with redshifted absorption, with local absorption,
and the broken power law with Galactic absorption,
we found the $\chi^2$ increments of $\Delta \chi^2= 4$, 12, and 30
for two additional d.o.f., 
and the calculated $F$=2.1, 6.4, 14.4, and 
the probability levels $P >5\%$, $<1\%$, $<1\%$, respectively.
Thus, regardless of the SIS calibration uncertainty, 
the second and the third models 
do not agree with the 0.5--0.7\,keV data at high confidence levels.
This can be understood as 
the redshifted absorption model, with its large 
absorption \nhe, predicts the strongest low-energy spectral cutoff,
i.e.\ the least count rates in 0.5--0.7\,keV.

It has been claimed that at 0.6\,keV the decrease of 
the efficiencies can be as much as 40\% for SIS1 in the current observations
(compared with GIS, ASCA Guest Observer Facility web page). 
We took this systematic effect into account by lowering the model values
in the 0.5--0.7\,keV bin by 40\% for (conservatively) both SIS0 and SIS1, 
and repeated the above testing.  
We have $\Delta \chi^2=$1, 5, and 9.5 
for the above three corresponding models, respectively.
This means while the first and the second models give the best and marginal
consistencies with the 0.5--0.7\,keV data ($F$=0.6 and 2.7), respectively,
the broken power law model still disagrees with the included data
at high confidence level ($F$=5, $P<1\%$).

Secondly, we tried to compensate the SIS calibration uncertainties
by decreasing the detecting efficiencies in the 0.5--0.7\,keV band  by
20\% for SIS0 and by 40\% for SIS1, 
and fitted the SIS+GIS spectra in the {\em whole} 0.5--9.5\,keV band
using the `corrected' SIS calibration files.
This resulted in the best-fit model being the
power law with excess absorption in the quasar frame,
with the goodness-of-fit  ($\chi^2$=190 for 212 d.o.f.)
improved slightly over the 
power law with local absorption model ($\chi^2$=193 for 212 d.o.f.),
and considerably over 
the broken power law with Galactic absorption model
($\chi^2$=197 for 211 d.o.f.).
The fitted spectral parameters are similar to those obtained
by fitting the 0.7--9.5\,keV spectra.
It is interesting to note that, for the excess absorption models,
the redshift of the absorber could be free fitted as 
$z=4.2^{+0.6}_{-0.5}$ (90\% c.l.);
however, this result is only suggestive given the
non-uniform efficiency loss over the 0.5--0.7\,keV band and
the less strict treatment here. 

We conclude that the broken power law with Galactic absorption model
gives a poor description to the observed low-energy spectrum down to 0.5\,keV,
in comparison with the (redshifted) excess absorption model.
Furthermore, the SIS data in the 0.5--0.7\,keV range
are better described by the absorption model with the absorber being at or close
to the quasar redshift than the local absorption model.

\section{Temporal properties}
The 67-hour duration of the observation corresponds to
$\sim 12.7$ hours in the quasar frame, 
which is short for quasar timing analysis.
The X-ray variability was searched for 
from the background-subtracted light curve
by using the packages provided in the FTOOLS. 
No statistically significant variations were found in both the 
co-added SIS and GIS light curves, 
as well as in the combined SIS+GIS light curve. 
It is noted, however, that a marginal, small amplitude variation 
could probably be present in the first half of the duration
($\chi^2$ test, $\chi^2=15$ for 9 d.o.f., a probability level of 0.09). 

We compared our results with the RASS flux 
by extrapolating the best-fit model down to 0.1\,keV. 
The predicted count rate in the ROSAT 0.1--2.4\,keV band is 
0.02\,cts\,s$^{-1}$,
consistent with the RASS measurement $0.035\pm0.011$\,cts\,s$^{-1}$
\citep{zickg97} within the errors.
Thus, no significant flux variation was found by comparing the two observations 
of about 2 years apart in the source rest frame.
It should be noted that, in the ROSAT soft 0.1--0.41\,keV band,
only $\sim 0.1$ source count is predicted
for the 424\,second RASS exposure due to the strong absorption of soft X-rays,
in contrast to the observed $\sim 2$ counts 
(a counting statistical probability of 3\% only).
This may imply the latter to be either of background fluctuations, 
or source photons scattered off 
the absorbing medium down to the soft X-ray energies.

\section{Discussion}
We have shown that the X-ray spectrum of \src exhibits a
low-energy cutoff much strong than that expected from the Galactic absorption.
Though both, the power law with excess absorption and 
the broken power law with Galactic absorption models give acceptable fits to
the spectra, the former is favored over the latter based on 
the fitting statistics. 
On the other hand, the power law model has been found to be a
characteristic of the intrinsic X-ray continuum emission of FSRQ 
over a wide range of redshift.
We thus suggest that the broken power law model seems to be an unlikely case,
though it cannot be ruled out confidently, and
the observed excess spectral cutoff is most likely caused by
absorption of the soft X-ray photons from the quasar.
This explanation is much natural in the light of 
what has been established in the redshift range of 
around 2 to 3---excess X-ray absorption is common in radio quasars,
and extends this behavior up to redshift above 4
(two out of the three objects with well determined X-ray spectra).
We consider this scenario only and discuss its implications in below.

\subsection{The absorber}
\label{dis:absorber}

The derived column density of the excess absorption 
(hereafter denoted simply as \nhe)
is among the largest\footnote{Similar 
amounts of excess absorption were also reported in 
S5\,0014+81 \citep{cappi97} and 
PKS\,0528+134 \citep{reeves}.}
ever found for high-$z$ radio quasars, 
which imposes a strong observational
constraint on the nature of the absorbing material. 
In addition, any satisfactory absorption models 
should be able to accommodate the absorber to the following facts: 
({\it a}) the relatively well constrained \nh--$z$ relation 
(Fig~\ref{fig:cont-nh-z}), though with some uncertainties 
introduced from the unknown metallicity;
({\it b}) being nearly neutral or moderately ionized, $\xi < 2$ at the 90\% c.l.
(unless the iron abundance relative to other heavy elements is abnormally 
sub-solar; see above); 
({\it c}) no strong absorption beyond the Lyman limit,
as indicated by both the optical (UV in the source frame) spectrum
and the measured $B$-magnitude (emitted at 834\,\AA) in \citet{zickg97}; 
and similarly,
({\it d}) no indication of heavy dust extinction;
({\it e}) if the X-ray emission is mainly from the jet 
(a likely case, see Sect.\,\ref{subs:qsonature}),
the absorber must lie outside the jet from the nucleus or around the jet.
Several possible origins of the X-ray absorption
of high-$z$ radio quasars have been proposed (see Elvis et al.\ 1994, 1998), 
such as damped Ly${\alpha}$ systems, intracluster gas, and the boundary layer
between the jet and surrounding medium.
We discuss in below the implications of our results 
for the possible absorbers of both intervening and intrinsic origins.

\subsubsection{Cosmological intervening material?}
\label{dis:absorb_galac}
An inspection of the optical image of the quasar field obtained by \citet{zickg97}
suggests readily an absorber candidate:
an extended, faint (speculative) spiral galaxy, 
which lies $\sim 7\arcsec$ away from the quasar
and seems not highly inclined given the estimated optical axes of 
$\sim 5\arcsec\times3\arcsec$. 
The presumed \hi gas disk/halo might intersect the quasar line-of-sight 
at a transverse distance of from a few kpc ($z\ll 1$) to several tens kpc
($z\sim 1$).
If this object is indeed a spiral galaxy, 
absorption of the quasar X-rays by its \hi gas is unavoidable.
The question is whether the amount of the absorbing gas can account for the
large column density derived here.
Observational evidence for X-ray absorption due to galactic \hi disk/shell
has just emerged in a few systems \citep{elvis97,yuan99},
but the measured \nh are relatively low (several times $10^{20}$\,\unhe).
We tried to assess quantitatively the amount of such possible absorption. 
We estimated the averaged
\nh of the anticipated \hi gas in this galaxy from its optical brightness,
by making use of the galactic \hi-mass and luminosity relation and 
the distance-independent scaling 
$N_{\rm H} \simeq 10^{20}(M_{\rm HI}/L_{\rm B}) 10^{0.4(27-\mu_{\rm B})}$
\citep{disney97},
where $M_{\rm HI}/L_{\rm B}$ is the \hi-mass to light ratio 
in units of $M_{\sun}/L_{\sun}$, 
and $\mu_{\rm B}$ the $B$-band surface brightness averaged over the \hi disk.
Adopting the measured $m_{\rm B}=20.2\pm0.5$,
we found a conservative $\mu_{\rm B}\sim 25.2$ (Galactic extinction corrected),
which is averaged over the solid angle confined 
by an ellipse with the inferred axis ratio 
and intersecting the quasar sight-line. 
This led to an averaged 
$N_{\rm H} \sim 5.2(^{+3.1}_{-1.9})~10^{20}(M_{\rm HI}/L_{\rm B})$\,\unhe,
where the quoted errors are from the uncertainty of $m_{\rm B}$. 
To match the excess absorption \nh of
$2.5(^{+0.5}_{-0.4})~10^{21}$\,\unh ($z=0$ and $A_{\rm Z}=A_{\rm Z \sun}$), 
a $M_{\rm HI}/L_{\rm B}$ ratio as high as 4--5 is needed;
this seems to be unlikely as spirals being found to have 
$M_{\rm HI}/L_{\rm B}\lesssim 1$ 
regardless of the morphological type (e.g.\ Kamphuis et al.\ 1996).
Moreover, the true \nh at the site of the quasar sight-line
might be even lower than the averaged value 
as the \hi surface density drops rapidly outward from the center.
We therefore suggest that this galaxy is probably responsible for
part of the X-ray absorption, 
but it is unlikely---though we cannot rule it out---to 
be able to account for the {\em total} absorption opacity.
After subtracting the plausible contribution of this galaxy 
(assuming $M_{\rm HI}/L_{\rm B}\la 1$),
an excess column density of \nh$\gtrsim 2.0(^{+0.5}_{-0.4})~10^{21}$\,\unh 
($z=0$ and $A_{\rm Z}=A_{\rm Z \sun}$) 
may remain unaccounted for.
 
Such a large amount of absorption seems not to be easily explained by
the general damped Ly${\alpha}$ systems (DLA),
neutral hydrogen reservoirs with the largest column density (\nhie)
known in the universe.
Although DLA may have the distribution of \nhi
reaching  a few times $10^{21}$\,\unh at its high end, 
the steep distribution function with a power law index of $-3$
renders such systems to be extremely rare 
(Wolfe et al.\ 1995, Zwaan et al.\ 1999).
The deficit of column density is even severe regarding 
the substantially sub-solar metal abundances ($A_{\rm Z}<0.1A_{\rm Z \sun}$)
of DLA, even at $z\simeq 0$ \citep{miller99}, 
which require \nh to be well above $10^{22}$\,\unh 
in order to produce the observed absorption.
Although the true total hydrogen column density \nh could be somewhat 
higher than \nhi considering a possible ionization fraction of H,
the correction should not be large as gases in DLA are
believed to be mainly neutral \citep{wolfe93}.
A detailed study by \citet{oflah97} incorporating absorber statistics
confirmed the extremely low detectability of high-\nh intervening systems;
the chance probability to have a DLA with \nh$\ga 10^{21}$\,\unh
on the line-of-sight to a high-$z$ quasar is less than a few percent.
This probability is even lower considering that any DLA systems, 
if do produce the absorption, are most likely
at low redshifts so as to avoid the otherwise 
much higher X-ray \nh required (Fig.\,\ref{fig:cont-nh-z}).
The non-detection of a DLA in the optical spectrum obtained by \citet{zickg97}
places an upper limit on the redshift of any possible DLA as $\sim 2.5$.
Therefore, the likelihood of the absorption being caused by 
cosmologically intervening material is regarded to be low.

\subsubsection{Intrinsic absorber?}
\label{dis:absorb_intri}
Alternatively, the excess absorption might take place 
intrinsically close to the quasar. 
This seems to be plausible in the light of the statistical arguments 
for high-$z$ radio-quiet quasars showing no excess absorption.
The inferred absorption column density in the quasar rest frame then
reaches the highest ever found among this type, 
$\sim 2~10^{23}$\,\unh 
($\ga 1.7~10^{23}$\,\unh corrected for the possible foreground
galactic absorption assuming the expected $M_{\rm HI}/L_{\rm B}\la 1$,
see Sect.\,\ref{dis:absorb_galac}) 
or even higher up to 
$\sim 10^{24}$\,\unh if $A_{\rm Z} \la 0.1 A_{\rm Z\sun}$.
Such thick gas can not be easily explained by the known mass components.
This result strengthens the tentative, apparent trend of the
increasing \nh of excess absorption in radio quasars 
with the increase of redshift, 
which might have interesting consequences for understanding
the evolution of quasars and, probably, 
the formation of the cosmic structure. 

One immediate problem,
which seems to be common among objects of this type \citep{elvis94},
arises as the apparent inconsistency between the X-ray and optical data: 
the thick gas, if contains dust similar to that of our Galaxy in
both composition and content, would have blocked almost all the 
optical/UV light of the quasar believed to originate from the central region, 
e.g.\ $A(\lambda 1230\mbox{\AA})$=40 \citep{seaton79} 
for the observed $R$-magnitude. 
Though this problem may be relieved by a likely extremely low dust content
or different dust compositions at high redshift 
(see Elvis et al.\ 1994 for a discussion), 
the lack of strong Lyman limit absorption remains a concern. 
A rough estimate of $\tau_{912\mbox{\AA}}\la 1$ 
from the optical spectrum in \citet{zickg97}
suggests the absorber to be highly ionized, \nhi/\nh $\la 10^{-6}$ 
(taking the conservative value of \nh=$1.7~10^{23}$\,\unhe).
This ionization state requires the dimensionless ionization parameter $U\ga 5$ 
($U$ is defined as the hydrogen-ionizing photon-to-electron number density ratio)
for the typical AGN ionizing continuum,
or, when converted to $\xi$, $\xi \ga 250$;
assuming the ionizing continuum as the extrapolation of the observed
flat power law with $\Gamma\simeq 1.7$ requires $\xi \ga 370$.
However, we found $\xi \la 2$ (20) at the 90\% (99\%) c.l. from the X-ray data.
One natural explanation would be that 
the UV continuum emitting region is only partially or not covered 
by the X-ray absorber.
If this is the case, special geometry of the absorbing gases may be required;
for instance, the gas might be ambiently local to the jet 
if the X-rays are of jet emission (constraint {\it e}).
Again, it should be noted that, 
if the gas has iron-poor metallicity in comparison to the solar one,
a higher ionization state cannot be ruled out.

Interestingly, \citet{elvis94} related the X-ray absorption 
of quasars up to $z\sim3$
to cooling flows of the possible host clusters of galaxies, 
though the existence of relaxed clusters at such high redshifts is unknown. 
If we take this hypothesis and extend the redshift up to $z\sim 4$ 
for the case of \srce, 
its large \nh value
in comparison to those found from X-ray absorption in
low-redshift clusters ($\sim 10^{21}$\,\unhe, White et al.\ 1991)
would imply an extremely strong cooling flow.
The mass of the cooled gas reaches 
$\sim 3~10^{13}R_{100}^2$\,M$_{\sun}$ 
for a conservative \nhe=$10^{23}$\,\unhe,
where $R_{100}$ is the cooling flow radius in units of 100\,kpc.
For $R_{100}\sim 1$ as typically found in low-$z$ clusters,  
this means a cooled mass of $\sim 3~10^{13}$\,M$_{\sun}$,
and a mass cooling rate $\gtrsim 3~10^{4}$\,M$_{\sun}$\,yr$^{-1}$
given the short life-time of the system of less than $10^{9}$ years. 
Such a cooling flow almost approaches the `maximal cooling flow' 
in which the cooling time is comparable 
with the gravitational free-fall time \citep{fabian94}.
Though these numbers will be somewhat reduced for a likely clumpy distribution
of the cooled gas, a cooling flow stronger than 
those in nearby clusters is suggested. 

It has also been noted that the massive ($\sim 10^{14}$\,M$_{\sun}$) 
neutral hydrogen cloud at $z$=3.4 reported by \citet{uson91} 
could be a possible origin of X-ray absorption.
The gas cloud was detected 
via both absorption and emission lines at the radio 21\,cm wavelength, 
and was interpreted as indications of a proto-cluster of galaxies. 
The \hi column density derived from the absorption line
is $4.4~10^{18}(T_{\rm s}/{\rm K})$\,\unh 
($T_{\rm s}$ the spin temperature of \hie);
the mean surface mass density was estimated from the emission line
as $\sim 48$\,M$_{\sun}$\,pc$^{-2}$ \citep{uson91},
which corresponds to an averaged column density 
\nhi$\sim 5.7~10^{21}$\,\unhe.
If such a system is responsible for the X-ray absorption in \srce,
the gas must be hot with $T_{\rm s}>10^4$\,K, and/or somewhat ionized,
\nh/\nhi$>10$; these lower limits will be further raised up for \abd$<$\abdsole.
Thus, the inferred $T_{\rm s}$ and/or total gas mass are likely exceeding
$10^4$\,K and $3 ~10^{14}$\,M$_{\sun}$, respectively, 
the values proposed for a ``Zel'dovich pancake" \citep{sunyaev}. 

If the quasar X-rays are dominated by jet emission,
the absorption may be caused by intimately jet-linked material.  
Such a scenario explains naturally the lack of UV absorption and 
reddening, which seems to be a commonplace for high-$z$ radio quasars showing
excess X-ray absorption. 
It is of particular interest to relate the X-ray absorption 
with the optical-radio `alignment effect' found in high-$z$ radio galaxies,
as noted in \citet{elvis98}.
This effect is the alignment of the extended, 
polarized optical emission with the radio axes (e.g.\ McCarthy 1993),
which is thought to be of (dust) scattering and/or star-formation origin.
A class of dust scattering models advocates the existence of gas-dust clouds
embedded in a multi-phase boundary layer formed (from interaction)
between the jet and ambient medium (e.g.\ De Young 1998). 
Although such a jet environment is largely uncertain,
it is interesting to note the following order-of-magnitude estimation.
We consider a specific set of model parameters 
which was adopted so as to roughly satisfy 
the physical condition and observations as proposed in \citet{deyoung98}:
gas density $n$=10--100\,cm$^{-3}$, cloud size $R_{\rm c}\sim$100\,pc.
We assume the boundary layer containing one layer of similar clouds, i.e.\ 
its height is characterized by the size of the clouds $R_{\rm c}$.
Consider an X-ray photon emitted from the jet   
with an angle $\theta$ with respect to the jet axis in the observer frame; 
for FSRQ, $\theta$ is small, $\theta\la 10^{\degr}$.
The average number of the clouds that the photon is expected to encounter 
before it escapes out of this `cocoon' is 
$N_{\rm enc}=R_{\rm c}/(\sin\theta\, d_{\rm c})$, 
where $d_{\rm c}$ is the mean distance between two neighboring clouds.
If the clouds are abundant enough, 
$d_{\rm c}/R_{\rm c} \la (\sin\theta)^{-1}$,
$N_{\rm enc}$ reaches unity ($N_{\rm enc}\ga 1$) and
results in X-ray absorption with almost a full coverage 
as viewed from the observer. 
The column density is then
\nh $\sim n R_{\rm c} N_{\rm enc} = 3 N_{\rm enc} ~10^{21-22}$\,\unhe.
Thus, for only one encounter on average ($N_{\rm enc}$=1)
the resulting \nh is sufficient to account for the X-ray column densities
found in most of the excess absorption quasars.
To reach \nh$\sim 10^{23}$\,\unh in \srce, an increase in $N_{\rm enc}$
up to  several encounters is required; 
this may be achieved by a rather small viewing angle to the jet $\theta$,
and/or an increased cloud number density (reduced $d_{\rm c}$).
For such a model, the presence of X-ray absorption in radio quasars
might be governed by the formation of such a boundary layer, 
for which cosmic evolution is needed to explain the difference at
high and low redshifts.

\subsection{The quasar}
\label{subs:qsonature}
The 2--50\,keV luminosity of $2.6~10^{47}$\,\ulume, 
after correction for the total absorption, 
makes \src one of the most luminous quasars. 
Furthermore, if $A_{\rm Z} \la 0.1$\,\abdsole, 
we have \nh$\ga 10^{24}$\,\unh and the Thomson depth 
$\tau_{\rm Th}\simeq 1.2 \sigma_{\rm Th} N_{\rm H}$ reaches unity
for photons with energies above a few keV 
(so that electrons bound to atoms can be treated as free); 
electron scattering becomes non-negligible and actually 
dominating the opacity for photons with $E\ga 5$\,keV.
An additional correction for such electron scattering opacity 
would raise the 2--50\,keV luminosity at least a few times higher, 
perhaps approaching $10^{48}$\,\ulume.
This will make the quasar one of the most luminous steady objects 
if the radiation is isotropic.

\begin{figure}
\psfig{figure=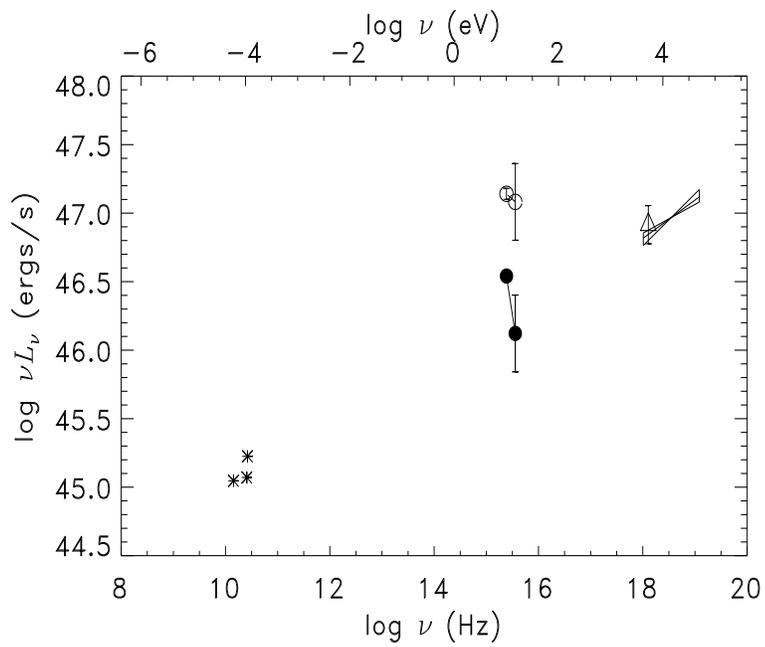,height=9.0cm,width=11.0cm,angle=0}
\caption[]{Spectral energy distribution of the quasar 
in the source rest frame, in which the 
luminosities are at the emitting frequencies. 
The X-ray luminosities are corrected for the total absorption.
The RASS measurement (triangle) is converted from the count rate assuming
the spectral model as the extrapolation of the best-fit absorbed power law 
derived in this work, and at a frequency corresponding to 1\,keV in observer
frame.   
The UV luminosities 
are obtained from the optical B- and R-band (non-simultaneously) photometry
data \citep{zickg97}, with the correction for dust extinction applied for:
(a) the Galactic extinction only (filled dots); 
(b) the Galactic plus an addition column of 
a local X-ray absorber ($z\lesssim 0.1$) assuming the 
Galactic dust-to-gas ratio (open circles).
In the case of the excess absorption occurring at a high redshift or 
intrinsic to the quasar, the {\it intrinsic} UV luminosities 
are uncertain due to the unknown extinction;
however, they must be no less than
the observed values for which only the 
correction for the Galactic reddening was applied (filled dots).  
}
\label{fig:sed}
\end{figure}

Fig.\,\ref{fig:sed} shows the spectral energy distribution (SED) for \srce,
with the X-ray luminosity corrected for the total absorption.
The UV luminosities ($L_{\rm UV}$) are somewhat uncertain due to 
the unknown dust extinction possibly imposed by the X-ray absorber. 
For a comparison, also plotted are $L_{\rm UV}$ (open circles)
for one extreme case in which the absorption
is solely attributed to a local intervening (galactic) absorber 
with the Galactic dust-to-gas ratio;
$L_{\rm UV}$ then reaches $10^{47}$\,\ulume. 
The situation is less clear for the case of intrinsic absorption,
as discussed above; 
in any case $L_{\rm UV}$ must be no less than
the observed values for which only the 
correction for the Galactic reddening was applied (filled dots).

It has been pointed out by \citet{fabian99} that this object
(and the other two $z>4$ quasars) is distributed somewhat apart from
the bulk of nearby blazars on the \alprx--\alpox and \alpro--\alpox planes,
where \alproe, \alprxe, and \alpox are the so-called broad band
effective spectral indices\footnote{Defined as
\alproe=$-\log(L_{\rm o}/L_{\rm r})/\log(\nu_{\rm o}/\nu_{\rm r})$,
\alprxe=$-\log(L_{\rm x}/L_{\rm r})/\log(\nu_{\rm x}/\nu_{\rm r})$, and
\alpoxe=$-\log(L_{\rm x}/L_{\rm o})/\log(\nu_{\rm x}/\nu_{\rm o})$,
where $L_{\rm r}$, $L_{\rm o}$, and $L_{\rm x}$ are 
the monochromatic luminosities at
the radio $\nu_{\rm r}$, optical $\nu_{\rm o}$, and X-ray $\nu_{\rm x}$
bands, respectively.
}
spanning between the radio, optical, and X-ray wavebands, respectively.
For \srce, this might be, at least partly, 
a consequence of the optical/UV extinction 
imposed by the obscuring matter. 
The intrinsic \alpro and \alpox would flatten and steepen, respectively,
and this would shift the object toward or into the bulk distribution
of FSRQ and radio-selected BL Lac objects in the \alpro-- \alprx-- \alpox 
parameter space (see Fig.\,4 in Fabian et al.\ 1999),
depending on the amount of dust extinction.   
For example, assuming local absorption and the dust-to-gas ratio similar
to the Galactic one, 
we found\footnote{The radio, optical and X-ray luminosities are calculated
at 5\,GHz, 5500\,\AA and 1keV, and an optical spectral index of $-0.7$ is
assumed for K-correction.} intrinsic \alproe=0.56, \alprxe=0.75, and \alpoxe=1.11.

Both, the broad band SED and the extreme X-ray luminosity suggest that
\src might be a blazar type object; 
its high radio power is also of typical FSRQ.
We may thus come to the suggestion that all these three $z>4$ quasars,
regardless of the strong X-ray absorption of \srce, are of the same class,
and the X-rays are thought to be emitted via inverse Compton radiation 
\citep{jones74}.
Similar to \gb as discussed in \citet{fabian99},
the optical/UV luminosity of \src
might not be dominated by the beamed non-thermal emission,
but originate from accretion process; 
if this is true, the high luminosity would
imply that black holes as massive as $10^{8}-10^{9}$\,M$_{\sun}$ have formed  
within less than 1 billion years in the early stages of the universe. 

Owing largely to its low-energy cutoff, the X-ray spectrum of the quasar
is consistent with that of the cosmic X-ray background (CXB)
in the 0.6--10\,keV ASCA band,
$\Gamma$=1.4 power law \citep{gendreau}
or $kT=$40\,keV bremsstrahlung (3--50\,keV, Marshall et al.\ 1980)
with Galactic absorption 
(fits with the parameters fixed at these values resulted in
reduced $\chi^2$=1.1 and 1.0 for 212 d.o.f., respectively).
However, this type of objects appears simply too rare to make significant
contribution to the CXB.

Finally, it is noted that the $z$=4.3 quasar 1508+5714
revealed a high rest-frame temperature $93(^{+40}_{-24})$\,keV (90\% errors,
Moran \& Helfand 1997)
when fitted with a redshifted bremsstrahlung with Galactic absorption model.
This value is similar to that found for \src of $132(_{-27}^{+41})$\,keV,
which apparently resulted from the low-energy cutoff in its spectrum.
Such a comparison suggests possible excess absorption in 1508+5714 too,
of which the ASCA observation might not be deep enough to allow a detection.  
The 90\% upper limit on the excess absorption \nh at the quasar redshift
was given as $1.3~10^{22}$\,\unh \citep{moran97}.

\section{Summary}
We present the broad band 3--50\,keV (source rest frame) X-ray observation
of the quasar \src at $z =4.28$.
The spectrum, with its unprecedented high quality among observations of 
objects at similar redshifts,
reveals an evident, substantial low-energy cutoff in excess of that
due to Galactic absorption.
The spectrum is best modeled by a power law 
with excess photoelectric absorption model, 
yet a broken power law with Galactic absorption cannot be ruled out.  
We suggest the excess absorption to be the most likely explanation,
considering the spectral data at the low-energy end.      
The column density of the absorber,
depending on its redshift and metallicity,
ranges from $2.5~10^{21}$\,\unh for local absorption up to
$2.1~10^{23}$\,\unh (solar metallicity) or
$1.6~10^{24}$\,\unh (10\% solar metallicity) 
for absorption at the quasar redshift.
It has to be noted there may be a systematic uncertainty 
in the determination of the column density, 
which tends to overestimate the absorption;
however, we suggest this effect to be small and negligible 
in comparison to the statistical uncertainties.
We attribute a possible part of the absorption opacity 
to a putative foreground spiral galaxy; 
the remaining absorption, most likely the majority part,
seems not to be easily explained by cosmological intervening systems.
Future observations with XMM and Chandra would be able to measure the
redshift and other parameters of the absorber by 
detecting possible iron absorption edges above 7\,keV.

The unabsorbed quasar continuum is well described by a simple power law
with $\Gamma=1.67^{+0.07}_{-0.04}$ 
extending up to 50\,keV in the source rest frame.
The absorption corrected luminosity reaches as high as $2.6~10^{47}$\,\ulume,
regardless of possible electron scattering effect.
No statistically significant X-ray flux variations were found 
during the relatively short observational interval,
neither by a comparison with the early ROSAT Survey observation.
Considering its extreme apparent luminosities in the X-ray and radio
bands, as well as the broad band SED, we suggest \src to be a blazar type object,
similar to the other two $z>4$ quasars observed with broad band X-ray 
spectroscopy.

\acknowledgments
We thank all the members of ASCA team for making 
the observations and data analysis possible.
We would also like to thank the referee for a careful reading of the
manuscript and the useful comments which helped to improve the paper.
W.Y. acknowledges the STA fellowship and hospitality at NASDA.
This research has made use of 
the NASA/IPAC Extragalactic Database (NED), which is operated by
the JPL, Caltech, under contract with the 
National Aeronautics and Space Administration.

\begin{deluxetable}{lllll}
\tabletypesize{\scriptsize}
\tablecaption{ASCA observation and data screening \label{tab:observation}}
\tablewidth{0pt}
\tablehead{
\colhead{Item} & \colhead{SIS0}   & \colhead{SIS1}   & \colhead{GIS2} & \colhead{GIS3}  
}
\startdata
Elevation angle\tablenotemark{a} &  $15^{\circ}$/$30^{\circ}$ & $15^{\circ}$/$25^{\circ}$ & $15^{\circ}$ & $15^{\circ}$ \\
C.O.R.\tablenotemark{b}   &  8    &   8   &  8     &   8     \\
Exposure (sec)           & 59061 & 60533 & 55085  & 55090   \\
Counts\tablenotemark{c}  & 2179 &  1831 & 1620   &  1807    \\
Background counts\tablenotemark{d} & 675 (1.14)& 606 (1.08) & 657 (2.0) & 786 (2.0)  \\
Net count rates\tablenotemark{e} & 2.54$\pm$0.10 & 2.02$\pm$0.09 & 1.74$\pm$0.08 & 1.85$\pm$0.09      \\
\enddata
\tablenotetext{a}{For SIS the second number refers to the bright Earth elevation angle.}
\tablenotetext{b}{Magnetic cutoff rigidity in GeV/c} 
\tablenotetext{c}{Overall counts from the source extracting region}
\tablenotetext{d}{Background counts normalized to the source extracting region,
which are the extracted background counts 
divided by the ratio of the background-to-source region areas (in bracket).}
\tablenotetext{e}{Background subtracted count rate in units of 
$10^{-2}\,\rm cts\,s^{-1}$ in the 0.6--10\,keV band for SIS and the
0.7--10\,keV band for GIS.} 
\end{deluxetable}

\begin{deluxetable}{llllr}
\tablecaption{Results of X-ray spectral fits \label{tab:spec_fits}}
\tablecolumns{5}
\tablewidth{0pt}
\tablehead{
\multicolumn{5}{c}{Power law + local neutral absorption}\\
\colhead{Detector} & \colhead{\nhe\tablenotemark{a}} & \colhead{$\Gamma$} & \colhead{norm.\tablenotemark{b}} & \colhead{$\chi^2$/dof}\\
}
\startdata
GIS      & 0.0459 (fix) & $1.51\pm0.06$ & $1.98\pm 0.12$ & 110/100 \\
SIS0+SIS1& 0.0459 (fix) & $1.24\pm0.04$ & $1.38\pm 0.06$ & 102/112 \\
SIS+GIS  & 0.0459 (fix) & $1.34\pm0.03$ & $1.69\pm 0.07$ & 221/211 \\
GIS   & $0.30_{-0.10}^{+0.12}$ & $1.76_{-0.07}^{+0.12}$ & $2.81_{-0.42}^{+0.51}$ & 104/99\\
SIS0+SIS1& $0.26_{-0.05}^{+0.06}$ & $1.56\pm0.09$ & $2.06_{-0.17}^{+0.24}$ & 84/111 \\
SIS+GIS  & $0.30_{-0.04}^{+0.05}$ & $1.67_{-0.04}^{+0.07}$ & $2.61_{-0.21}^{+0.24}$ & 187/210\\
\cline{1-5}\\
\multicolumn{5}{c}{Power law + excess absorption at $z=4.28$ and \gnh (SIS+GIS)}\\
\colhead{\abde/\abdsol} & \colhead{\nhex} & \colhead{$\Gamma$} & \colhead{norm.} & \colhead{$\chi^2$/dof} \\ 
\cline{1-5}\\
1.0 (fix)& $21.1_{-4.2}^{+5.3}$ & $1.72_{-0.07}^{+0.10}$ & $2.86_{-0.26}^{+0.40}$ & 187/210 \\
0.1 (fix)  & $156_{-26}^{+31}$  & $1.72_{-0.08}^{+0.04}$ & $2.81\pm 0.50$ & 186/210 \\
\cline{1-5}\\
\multicolumn{5}{c}{Bremsstrahlung + local neutral absorption (SIS+GIS)}\\
\colhead{redshift} & \colhead{\nh}  &  \colhead{T (keV)} & \colhead{norm.\tablenotemark{c}}& \colhead{$\chi^2$/dof} \\
\cline{1-5}\\
4.276 (fix) & 0.0459 (fix) & $131.7_{-26.7}^{+40.5}$ & $3.53\pm0.16$ & 210/211 \\
4.276 (fix) & $0.21_{-0.03}^{+0.04}$ & $54.7_{-9.2}^{+12.2}$ & $3.75_{-0.16}^{+0.14}$ & 187/210 \\
\cline{1-5}\\
\multicolumn{5}{c}{Broken power law + local neutral absorption (SIS+GIS) }\\
\colhead{$E_{\rm break}$\tablenotemark{d}} & \colhead{\nh} & \colhead{$\Gamma_{\rm low}$} & \colhead{$\Gamma_{\rm high}$} & \colhead{$\chi^2$/dof}\\
\cline{1-5}\\
$1.7^{+0.2}_{-0.3}$ & 0.0459 (fix) & $0.70^{+0.18}_{-0.27}$ & $1.58^{+0.08}_{-0.11}$ & 190/209 \\
$2.0^{+\inf}_{-\inf}$ & $0.28\pm0.07$  &  $1.58^{+2.48}_{-0.38}$ & $1.67^{+0.34}_{-0.10}$ & 187/208 \\
\enddata
\tablenotetext{a}{Column density of hydrogen in units of $10^{22}$\,\unh}
\tablenotetext{b}{Nomalization in units of 
$10^{-4}$\,photons\,s$^{-1}$\,cm$^{-2}$\,keV$^{-1}$ at 1\,keV.
For joint SIS+GIS fits, the values given are of GIS, and
those of SIS are less by within 10\% in general.}
\tablenotetext{c}{Nomalization as given in {\rm XSPEC} in units of 
$10^{-14}/(4\pi D^2) \int n_{\rm e} n_{\rm H} dV$, 
where $D$ is the luminosity distance to the source (cm) and 
$n_{\rm e}$ and $n_{\rm H}$ the electron and hydrogen densities (cm$^{-3}$), respectively.} 
\tablenotetext{d}{The break energy of broken power law in keV. 
Un-specified uncertainties mean the value is unconstrained.}
\end{deluxetable}

\begin{thebibliography}{}
\bibitem[Barr et al.(1988)]{barr88} Barr, P., Giommi, P., Maccagni, D., 1988, \apjl, 324, 11 
\bibitem[Bechtold et al.(1994)]{bechtold94} Bechtold, J., Elvis, M., Fiore, F., et al., 1994, \aj, 108, 759
\bibitem[Boller et al.(2000)]{boller00} Boller, Th., Fabian, A. C., Brandt, W. N., and Freyberg, M. J., 2000, \mnras, in press
\bibitem[Brinkmann et al.(1997)]{brinkmann97} Brinkmann, W., Yuan, W., Siebert, J., 1997, \aap, 319, 413
\bibitem[Cappi et al.(1997)]{cappi97} Cappi, M., Matsuoka, M., Comastri, A., et al., 1997, \apj, 478, 492
\bibitem[De Young(1998)]{deyoung98} De Young, D. S., 1998, \apj, 507, 161
\bibitem[Dickey and Lockman(1990)]{dklckm} Dickey, J. M., \& Lockman, F. J. 1990, ARA\&A 28, 215
\bibitem[Disney and Banks(1997)]{disney97} Disney, M. J., \& Banks, G., 1997, PASA, 14, 69 
\bibitem[Done et al.(1992)]{done92} Done, C., Mulchaey, J. S., Mushotzky, R. F., Arnaud, K. A.,  1992, \apj, 395, 275
\bibitem[Dotani et al.(1996)]{dotani} Dotani, T., et al., 1996, ASCA News Letter, 4, 3
\bibitem[Elvis et al.(1994)]{elvis94} Elvis, M., Fiore, F., Wilkes, B., McDowell, J., Bechtold, J., 1994, \apj, 422, 60
\bibitem[Elvis et al.(1997)]{elvis97} Elvis, M., Fiore, F.,  Giommi, P., Padovani, P., 1997, \mnras, 291, L49
\bibitem[Elvis et al.(1998)]{elvis98} Elvis, M., Fiore, F.,  Giommi, P., Padovani, P., 1998, \apj, 492, 91
\bibitem[Fabian (1994)]{fabian94} Fabian, A. C., 1994, ARA\&A, 32, 277
\bibitem[Fabian et al.(1998)]{fabian98} Fabian, A. C., Iwasawa, K., Celotti, A., et al., 1998, \mnras, 295, L25
\bibitem[Fabian et al.(1999)]{fabian99} Fabian, A. C., Celotti, A., Pooley, G., et al., 1999, \mnras, 308, L6
\bibitem[Fiore et al.(1998)]{fiore98} Fiore, F., Elvis, M., Giommi, P., Padovani, P., 1998, \apj, 492, 79
\bibitem[Gendreau et al.(1995)]{gendreau} Gendreau, K. C., Mushotzky, R. F., Fabian, A. C., 1995, PASJ, 47, L5 
\bibitem[Jones et al.(1974)]{jones74} Jones, T. W., O'Dell, S. L., Stein, W. A., 1974, \apj, 188, 353
\bibitem[Kamphuis et al.(1996)]{kamphuis} Kamphuis, J. J., Sijbring, D., van Albada, T. S., 1996, \aaps, 116, 15
\bibitem[Kubo et al.(1997)]{kubo97} Kubo, H., Makishima, K., Takahashi, T., 1997, \mnras, 287, 328
\bibitem[Lawson and Turner(1997)]{lawson97} Lawson, A. J. and Turner, M. J. L., 1997, \mnras, 288, 920
\bibitem[Lu et al.(1996)]{lu96} Lu, L., Sargent, W.L.W., Barlow, T. A., Churchill, C. W., Vogt, S., 1996, \apjs, 107, 475
\bibitem[Madejski et al.(1991)]{madejski} Madejski, G. M., Mushotzky, R. F., Weaver K. A., Arnaud, K. A., Urry, C. M., 1991, \apj, 370, 198
\bibitem[Marshall et al.(1980)]{marshall80} Marshall, F. E., Boldt, E. A., Holt, S. S., et al., 1980, \apj, 235, 4
\bibitem[Matsuoka et al.(1999)]{matsuoka99} Matsuoka, M., Wang, T., Kubo, H., et al., 1999, in Highlights in X-ray Astronomy, ed., Aschenbach, B., and Freyberg, M., (MPE Report 272), 236
\bibitem[McCarthy (1993)]{mccarthy} McCarthy, P. J., 1993, ARA\&A, 31, 639
\bibitem[Miller et al.(1999)]{miller99} Miller, E. D., et al., 1999, \apjl, 510, L95
\bibitem[Moran and Helfand(1997)]{moran97} Moran, E. C., and Helfand D. J., 1997, \apjl, 484, L95
\bibitem[Morrison and McCammon(1983)]{wabs} Morrison, R., and McCammon, D., 1983, \apj, 270, 119
\bibitem[O'Flaherty and Jakobsen(1997)]{oflah97} O'Flaherty, K. S. and Jakobsen, P., 1997, \apj, 479, 673 
\bibitem[Otrupcek and Wright(1991)]{parkes} Otrupcek, R. E. and Wright, A. E., 1991, PASA, 9, 170
\bibitem[Pettini et al.(1997)]{pettini97} Pettini, M., Smith, L. J., King, D. L., Hunstead, R. W., 1997, \apj, 486, 665
\bibitem[Prochaska and Wolfe(2000)]{prochaska} Prochaska J. X. and Wolfe, A. W., 2000, astro-ph/0002513
\bibitem[Reeves et al.(1997)]{reeves} Reeves, J. N., Turner, M. J. L., Ohashi, T., Kii, T., 1997, \mnras, 292, 468
\bibitem[Schartel et al.(1997)]{schartel97} Schartel, N., Komossa, St., Brinkmann, W., et al., 1997, \aap, 320, 421
\bibitem[Seaton (1979)]{seaton79} Seaton, M. J., 1979, \mnras, 187, 73
\bibitem[Serlemitsos et al.(1994)]{serlem94} Serlemitsos, P., Yaqoob, T., Ricker, G., et al., 1994, \pasj, 46, L43
\bibitem[Siebert et al.(1996)]{siebert96} Siebert, J., Matsuoka, M., Brinkmann, W., et al., 1996, \aap, 307, 8
\bibitem[Sunyaev and Zel'dovich(1975)]{sunyaev} Sunyaev, R. A. and Zel'dovich, Ya. B., 1975, \mnras, 171, 375
\bibitem[Tanaka et al.(1994)]{tanaka94} Tanaka, Y., Inoue, H., Holt, S. S., 1994, \pasj, 46, L37 
\bibitem[Uson et al.(1991)]{uson91} Uson, J. M., Bagri, D. S., Cornwell, T. J., 1991, Phy. Rev. Lett., 67, 3328
\bibitem[White et al.(1994)]{white94} White, D. A., Fabian, A. C., Johnstone, R. M., Mushotzky, R. F., Arnaud, K. A., 1991, \mnras, 252, 72
\bibitem[Wilkes et al.(1992)]{wilkes92} Wilkes, B. J., Elvis, M., Fiore, F., et al., 1992, \apjl, 393, L1
\bibitem[Wolfe(1993)]{wolfe93} Wolfe, A. M., 1993, in Relativistic Astrophysics and Particle Cosmology, ed.\ C. W. Akerlof \& M. A. Srednicki (New York; Academy of Sciences), 281  
\bibitem[Wolfe et al.(1995)]{wolfe95} Wolfe, A. M., Lanzetta, K. M., Foltz, C. B., Chaffee F. H., 1995, \apj, 454, 698
\bibitem[Yaqoob et al.(2000)]{yaqoob00} Yaqoob, T., \& ASCA team, 2000, ASCA GOF Calibration Memo (ASCA-CAL-00-06-01, v1.0)
\bibitem[Yuan and Brinkmann(1999)]{yuan99} Yuan, W. and Brinkmann, W., 1999, in Highlights in X-ray Astronomy, ed., Aschenbach, B., and Freyberg, M., (MPE Report 272), 240  
\bibitem[Zickgraf et al.(1997)]{zickg97} Zickgraf, F.-J., Voges, W., Krautter, J., et al., 1997, \aap, 323, L21 
\bibitem[Zwaan et al.(1999)]{zwaan99} Zwaan, M. A., Verheijen M. A. W., Briggs, F. H., 1999, PASA, 16, 100
\end{thebibliography}
\end{document}